\documentclass{jpconf}

\pdfoutput=1

\usepackage[pdftex]{graphicx}
\usepackage{amssymb}
\usepackage{amsmath}
\usepackage{subfig}

\usepackage{xspace}
\xspaceaddexceptions{]}
\xspaceaddexceptions{"}

\newcommand{\figr}[1]{Figure~\ref{fig:#1}\xspace}
\newcommand{\sect}[1]{Section~\ref{sec:#1}\xspace}
\newcommand{\tbl}[1]{Table~\ref{tab:#1}\xspace}

\newcommand{\psra}{PSR~J0737$-$3039A\xspace}
\newcommand{\psrb}{PSR~J0737$-$3039B\xspace}

\newcommand{\fermi}{\textit{Fermi}~LAT\xspace}

\newcommand{\dg}{\ensuremath{^\circ}\xspace}
\newcommand{\rtg}{radio-to-\ensuremath{\gamma} phase lag\xspace}
\newcommand{\al}{\ensuremath{\alpha}\xspace}

\newcommand{\z}{\ensuremath{\zeta}\xspace}
\newcommand{\gr}{\ensuremath{\gamma}-ray\xspace}
\newcommand{\grs}{\ensuremath{\gamma}-rays\xspace}
\newcommand{\ppdot}{\ensuremath{P\text{-}\dot{P}\text{ diagram}}\xspace}
\newcommand{\rmax}{\ensuremath{R_{\rm max}}\xspace}
\newcommand{\rlc}{\ensuremath{R_{\rm LC}}\xspace}

\begin{document}
  \title{Modelling the $\gamma$-ray and radio light curves of the double pulsar system}

  \author{A S Seyffert$^1$, C Venter$^1$, A K Harding$^2$ and T J Johnson$^{3,4}$}
  \address{$^1$ Centre for Space Research, North-West University, Potchefstroom Campus, Private Bag X6001, Potchefstroom 2520, South Africa}
  \address{$^2$ Astrophysics Science Division, NASA Goddard Space Flight Center, Greenbelt, MD 20771, USA}
  \address{$^3$ National Research Council Research Associate}
  \address{$^4$ High-Energy Space Environment Branch, Naval Research Laboratory, Washington, DC 20375, USA}
  \ead{20126999@nwu.ac.za}

  \begin{abstract}
    Guillemot et al. recently reported the discovery of \gr pulsations from the 22.7~ms pulsar (pulsar A) in the famous double pulsar system J0737-3039A/B. The \gr light curve (LC) of pulsar A has two peaks separated by approximately half a rotation, and these are non-coincident with the observed radio and X-ray peaks. This suggests that the \gr emission originates in a part of the magnetosphere distinct from where the radio and X-ray radiation is generated. Thus far, three different methods have been applied to constrain the viewing geometry of pulsar A (its inclination and observer angles \al and \z): geometric modelling of the radio and \gr light curves, modelling of the position angle sweep in phase seen in the radio polarisation data, and independent studies of the time evolution of the radio pulse profile of pulsar A. These three independent, complementary methods have yielded consistent results: pulsar A's rotation axis is likely perpendicular to the orbital plane of the binary system, and its magnetic axis close to lying in the orbital plane (making this pulsar an orthogonal rotator). The observer is furthermore observing emission close to the magnetic axis. Thus far, however, current models could not reproduce all the characteristics of the radio and \gr light curves, specifically the large \rtg. In this paper we discuss some preliminary modelling attempts to address this problem, and offer ideas on how the LC fits may be improved by adapting the standard geometric models in order to reproduce the profile positions more accurately.
  \end{abstract}

  \section{Introduction}
    Using the Parkes multibeam receiver Burgay et al. reported the discovery of a 22\,ms pulsar, \psra, in a close binary system. The derived orbital parameters implied that the system consists of two neutron stars. \cite{Burgay03}, the short orbital period (roughly 2.4\,hr) coupled with the high orbital eccentricity (0.88) made this system the most rapidly merging neutron star binary yet discovered, with the eventual merger predicted to occur in approximately 85\,Myr. This discovery was soon followed by the discovery of radio pulsations from pulsar A's binary companion, \psrb, with a pulsation period of 2.8\,s \cite{Lyne04}. The fact that both stars are observed as radio pulsars, a kind of system that had not yet been found, further added to this system's uniqueness and allowed very sensitive, and famous, confirmation the predictions of Einstein's General Relativity using high precision measurements of the orbital motions of the two stars \cite{Kramer06}.

\begin{figure}[t]
      \centering
      \includegraphics[width=0.8\textwidth]{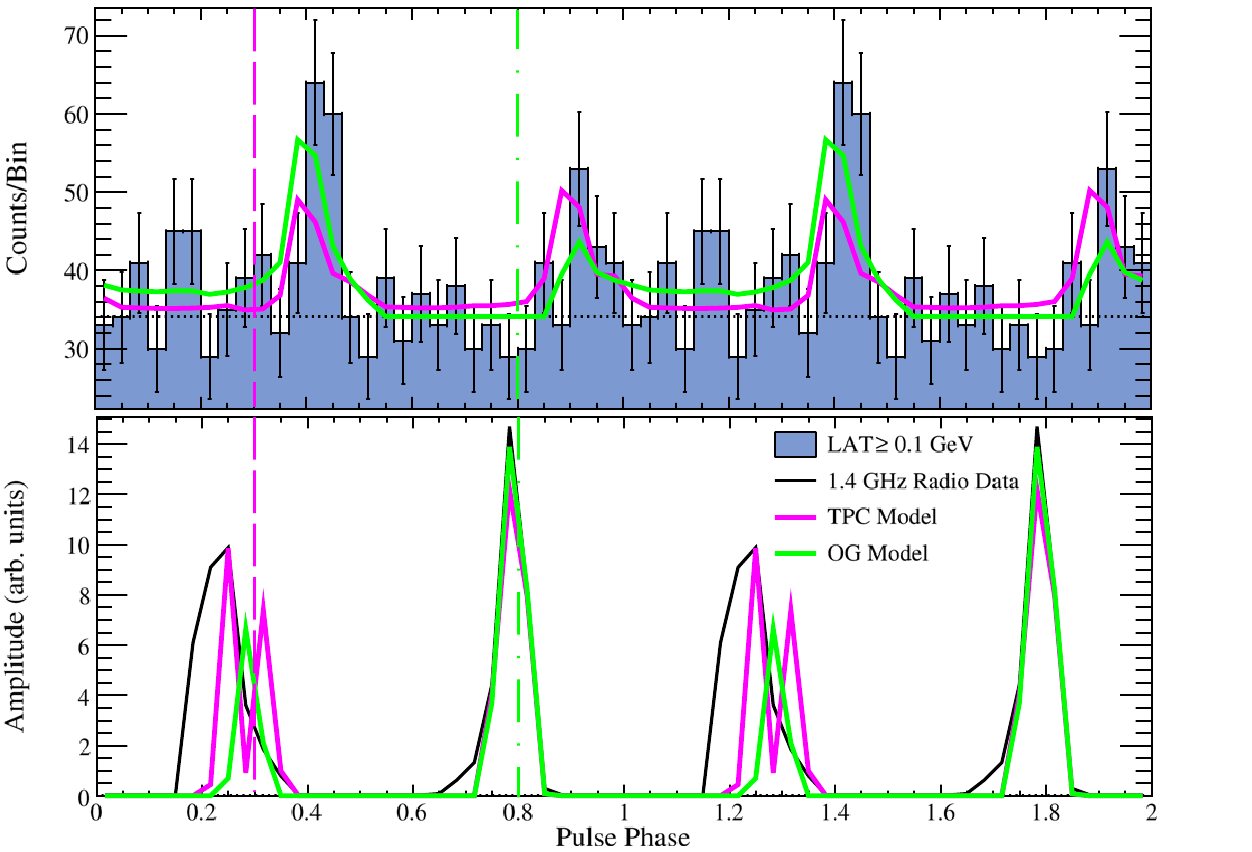}
      \caption{The observed radio (bottom) and \gr (top) LCs for \psra with best-fit solutions for the OG (green) and TPC (pink) geometric models. The vertical dashed and dash-dotted lines indicate the closest approach to the magnetic axis under the best-fit OG and TPC models respectively \cite{Guillemot13}.\label{fig:bestFits}}
    \end{figure}

    The discovery of pulsed \gr radiation from the millisecond pulsar \psra~\cite{Guillemot13} further added to the value of this system as \psra occupies a region on the \ppdot relatively devoid of \gr pulsars. \psra is the first mildly recycled millisecond pulsar yet observed to emit in \grs. \figr{bestFits} shows the radio and \gr light curves (LCs) observed for \psra at 1.4\,GHz and $\geqslant$0.1\,GeV respectively. Both LCs display a widely-spaced two-peak structure. Note that two rotations of the pulsar are shown in the interest of clarity.

  \section{Geometry}
    The first constraint on the viewing geometry of \psra that can be obtained from these LCs is that the two radio peaks are most likely associated with opposite magnetic poles as they are separated by about half a rotation in phase. This means that the inclination angle \al between the magnetic and rotation axes of this pulsar is most likely close to 90\dg. Furthermore, the large \rtg suggests that the radio and \gr radiation are produced in different regions of the magnetosphere. This large \rtg is, however, very troublesome when trying to reproduce these LCs through model simulations.

    \figr{bestFits} shows the best-fit LCs obtained for \psra using the outer gap (OG \cite{Cheng86_I,Cheng86_II}) and two-pole caustic (TPC \cite{Dyks03}) models for the \gr emission alongside an empirical conal model \cite{Story07} for the radio emission \cite{Guillemot13}. As can be seen in the bottom panel, the biggest difficulty when trying to fit these LCs is reproducing the large \rtg, with the leading radio peak in the predicted LCs still lying too close to the trailing peak.

    In addition to the radio and \gr LCs there are also high quality radio polarisation data available for \psra with which the viewing geometry can be constrained. Fitting these data using a modified rotating vector model \cite{Craig12} yields a geometry consistent with the one found using the geometric models, with \psra being an almost orthogonal rotator (see \tbl{table1}).

    The last estimate of the geometric parameters to mention here is the one obtained by Ferdman et al., who studied six years of radio observations of \psra \cite{Ferdman13}. The results of their study agree very well with the results of the other two approaches discussed above.

  \section{Improving the geometric fits\label{sec:improving}}
    \begin{figure}[t]
      \centering
      \subfloat[$\rmax=0.15\rlc$]{\includegraphics[width=0.5\textwidth]{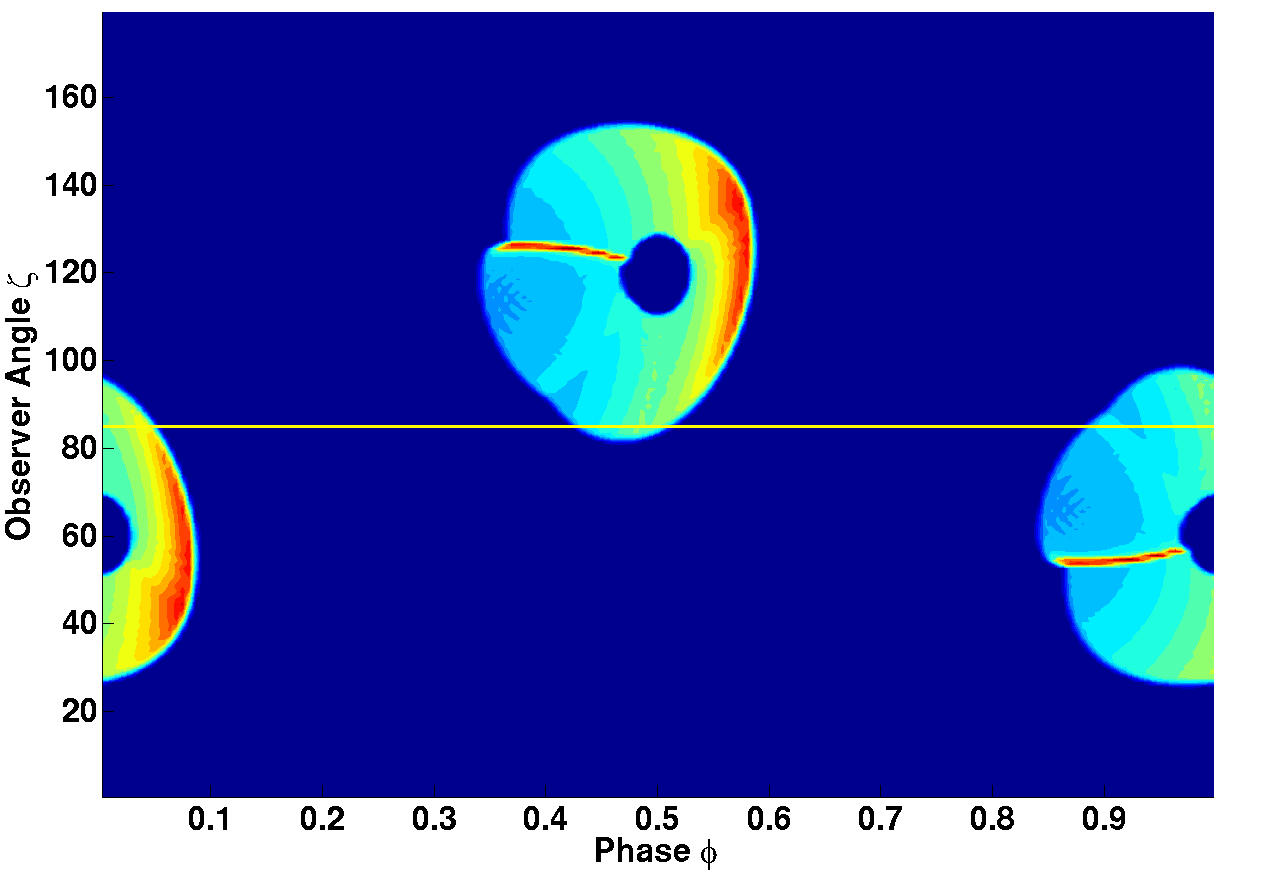}}
      \subfloat[$\rmax=0.20\rlc$]{\includegraphics[width=0.5\textwidth]{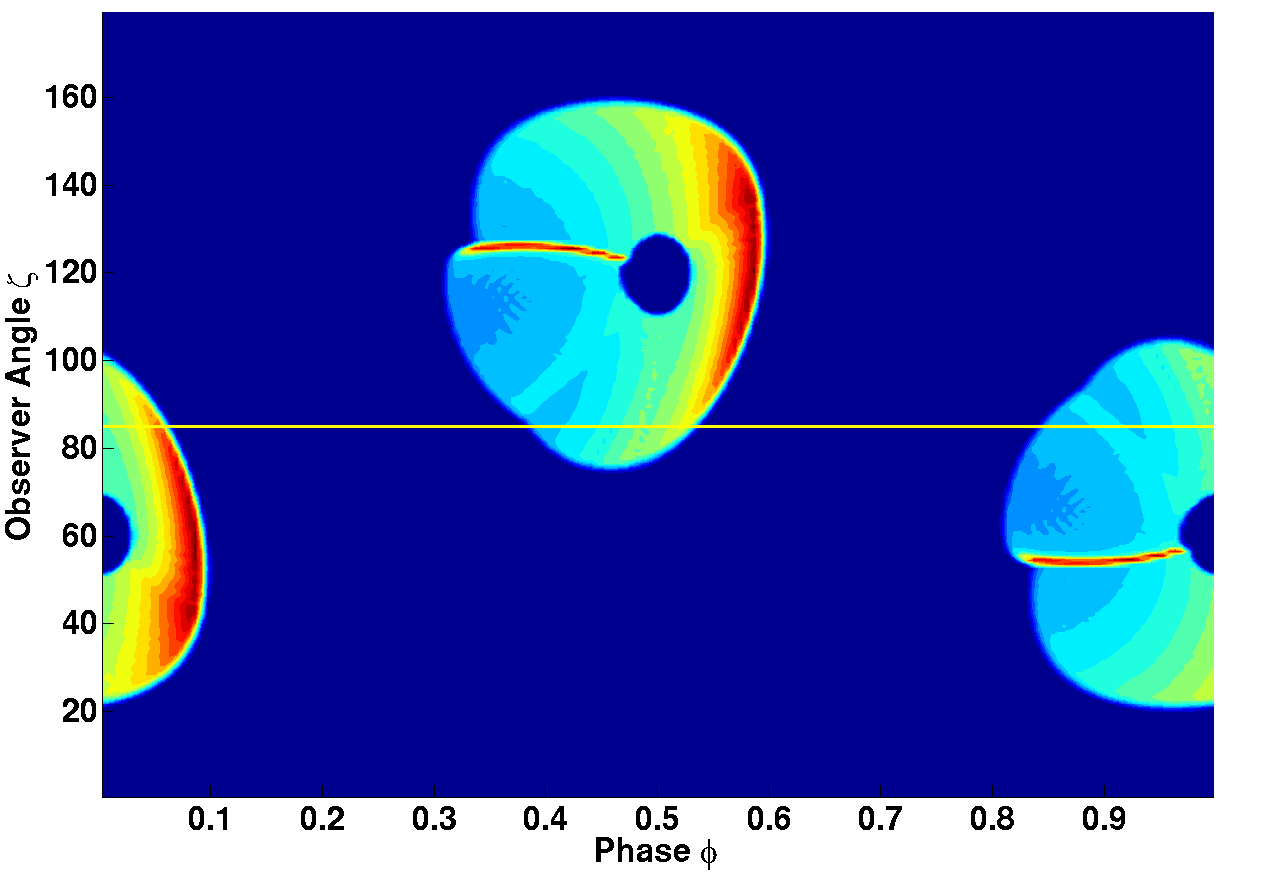}}\\
      \includegraphics[width=0.9\textwidth]{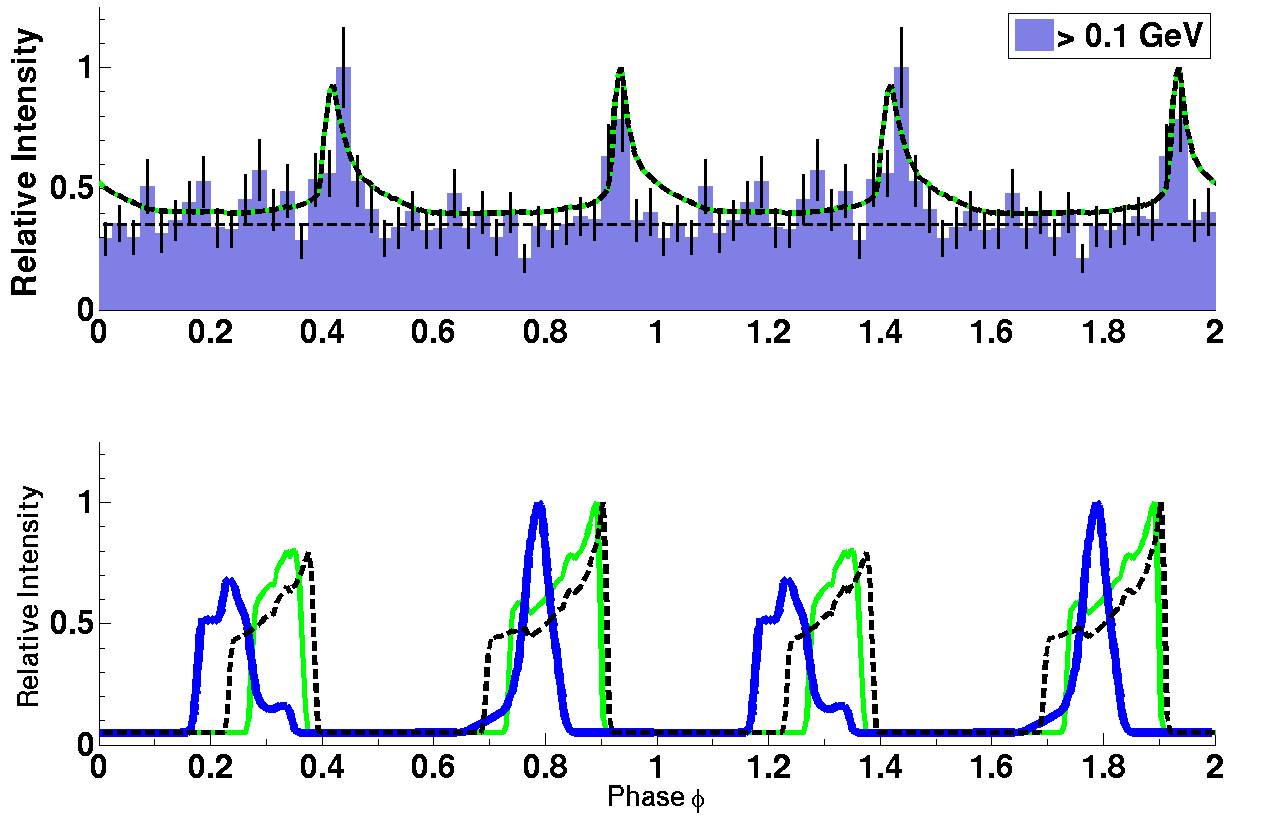}
      \caption{Fits obtained using the alternative model discussed in \sect{improving}. The green radio LC corresponds to a slot gap extending to a radius $\rmax = 0.15\rlc$, where \rlc is the light cylinder radius, and the dashed LC corresponds to $\rmax = 0.20\rlc$. In both cases the large \rtg cannot be reproduced. For illustration, we used $\al=60\dg$ and $\z=85\dg$.\label{fig:altModel1}}
    \end{figure}

    \begin{table}[b]
      \caption{\label{tab:table1}The geometric parameters of \psra as determined by three independent methods.}
      \begin{center}
        \begin{tabular}{llll}
          \br
          Method                    & \al (\dg) & \z (\dg) & Source\\
          \mr
          Geometic modelling        & \;\;\;$80^{+9}_{-3}$   & \;\;\;$86^{+2}_{-14}$  &\cite{Guillemot13}\\
          Polarisation data fitting & $98.8^{+8}_{-1.5}$     & $95.8^{+13.2}_{-4.3}$  &\cite{Craig12}\\
          Radio analysis            & $90.2^{+16.3}_{-16.2}$ & $90.8^{+0.27}_{-0.46}$ &\cite{Ferdman13}\\
          \br
        \end{tabular}
      \end{center}
    \end{table}

    The good agreement between the results yielded by the three independent approaches, as is summarized in \tbl{table1}, increases our confidence in the identification of \psra as an orthogonal rotator, but simultaneously poses a challenge to the models employed. Neither the OG nor TPC \gr models, when coupled with the conal radio model, are able to reproduce the large \rtg. A refinement of the current models is clearly necessary, and we have already made some attempts to rectify the problem.

    The first alternative model invoked a low-altitude slot gap geometry for the radio emission (see \cite{Arons83} for the original slot gap model), coupled with the usual TPC model for the \grs. This geometry is proposed in the context of a radio cone producing radio peaks that lead the caustic \gr peaks. The profiles produced by this model still could not reproduce the \rtg satisfactorily, even though the LC shapes were still reasonable. \figr{altModel1} shows fits obtained using this alternative model.

    The second alternative model proposed, instead, a conal structure for both the radio and \gr emitting regions, with the \gr region lying \textit{lower} than the radio region. This configuration was motivated by the idea that the radio may indeed have a dominating leading peak, with the \gr LC leading the radio LC. It was however found that it is not possible to reproduce both the radio and \gr profile shapes simultaneously within the context of this model, leading us to abandon this scenario.

  \section{Future work}
    The fits obtained using these preliminary alternate models may not be satisfactory, but they do point to interesting avenues of model refinement, e.g., the investigation of non-uniform emissivities, such as patchy or one-sided radio cones, or non-aligned radio and \gr cones. The fact that these LCs are hard to fit using the established geometric models, coupled with the unique characteristics of the system within which this pulsar finds itself, suggests that there may be some form of interaction between the two pulsars, perhaps through their stellar winds. Such an interaction may be observable through changes in the LCs at the orbital period, but thus far no such periodic phenomenon has been observed. It may also be the case that the stellar winds, and specifically the currents they constitute, perturb the usual magnetic field structure of the pulsars, and hence the geometry of the emitting regions. Such perturbations are currently not included in the geometric models employed thus far, and an investigation into how these two pulsars interact may lead to valuable refinements to the current geometric models.

  \ack CV is supported by the South African National Research Foundation. AKH acknowledges support from the NASA Astrophysics Theory Program. CV, TJJ, and AKH acknowledge support from the \textit{Fermi} Guest Investigator Program. The \fermi Collaboration acknowledges support from a number of agencies and institutes for both development and the operation of the LAT as well as scientific data analysis. These include NASA and DOE in the United States, CEA/Irfu and IN2P3/CNRS in France, ASI and INFN in Italy, MEXT, KEK, and JAXA in Japan, and the K.~A.~Wallenberg Foundation, the Swedish Research Council, and the National Space Board in Sweden. Additional support from INAF in Italy and CNES in France for science analysis during the operations phase is also gratefully acknowledged.

\end{document}